\RequirePackage{fix-cm}
\documentclass[twocolumn]{svjour3}          
\smartqed  

\usepackage[T1]{fontenc} 
\usepackage{xcolor}
\usepackage{graphicx}
\usepackage{dcolumn}
\usepackage{bm}
\usepackage{amsmath}
\usepackage{amssymb}
\usepackage{hyperref}
\usepackage[defaultcolor=red]{changes} 

\journalname{Granular Matter}

\begin{document}

\title{Average outpouring velocity and flow rate of grains discharged from a tilted quasi-2D silo
}

\author{ Ryan Kozlowski \and
        J. Carter Luketich \and
        Elijah Oshatz \and
        Douglas J. Durian \and
        Luis A. Pugnaloni 
}

\institute{R. Kozlowski \at
              Physics Department, College of the Holy Cross, Worcester, MA, 01610, USA \\
              \email{rkozlows@holycross.edu}   
           \and
           J. Carter Luketich \at
              Physics Department, Berea College, Berea, KY, 40404, USA
           \and
           Elijah Oshatz \at
              Physics Department, Berea College, Berea, KY, 40404, USA
           \and
           D. J. Durian \at
              Department of Physics and Astronomy, University of Pennsylvania, Philadelphia PA, 19104, USA
           \and
           L. A. Pugnaloni \at
              Departamento de F{\'i}sica, Facultad de Ciencias Exactas y Naturales, Universidad Nacional de La Pampa, CONICET, Uruguay 151, 6300 Santa Rosa (La Pampa), Argentina
}

\date{Received: date / Accepted: date}

\maketitle

\begin{abstract}
The flow of granular materials through constricted openings is important in many natural and industrial processes. These complex flows ---featuring dense, dissipative flow in the bulk but low-dissipation, low density outpouring in the vicinity of the orifice--- have long been characterized empirically by the Beverloo rule and, recently, modeled successfully using energy balance. The dependence of flow rate on the silo's angle with respect to gravity, however, is not captured by current models. We experimentally investigate the role of tilt angle in this work using a quasi-2D monolayer of grains in a silo. We measure mass flow rate, the average exit velocities of grains, and the packing fraction along the orifice with varying tilt angles. We propose a model that describes our results (and earlier findings with 3D systems [H. G. Sheldon and D. J. Durian, Granul. Matter 12, 579 (2010)]) by considering the dependence of outpouring speed and angle with respect to the orifice angle and, importantly, the angle of stagnant zones adjacent to the orifice. We conclude by posing questions about possible extensions of our model in order to describe spatial variations of exit velocity and density along the orifice cross section.
\end{abstract}

\section{Introduction\label{sec:introduction}}

Numerous systems in nature are characterized by the flow of particles through constricted openings, such as sand flowing in an hourglass, suspensions in microfluidic channels~\cite{Weeks2012HopperFlowEmulsions,Dessaire2017CloggingMicrofluidicsReview,Higgins2007SickleCellMicrofluidic}, and even pedestrians passing through a door~\cite{Helbing2000EscapePanic}. The flow and clogging of granular materials, in particular, is of broad interest for both applied (\textit{e.g.}, pharmaceutical)~\cite{Muzzio2002PowderTechnology}, geophysical~\cite{Liu2019MigrationCloggingPourousMedia}, and fundamental scientific reasons~\cite{Duran2000BookSPG,Jaeger1996GMSolidLiquidGas,Zuriguel2014BottleneckCloggingReview}. The flow of grains through a constricted orifice in a silo has long been known to follow the empirical Beverloo rule~\cite{Beverloo1961FlowRate,Mankoc2007FlowRateGMSilo}, given by 
\begin{equation*}
    Q = C\rho_b \sqrt{g}(D-kd)^{f},
\end{equation*}
where $f=3/2$ in two-dimensional (2D) systems and $f=5/2$ in three-dimensional (3D) systems, $C$ is a dimensionless fit parameter that is largely independent of details of the granular material~\cite{Nedderman1992StaticsAndKinematicsGM,Darias2020EnergyBalanceBeverloo}, $D$ is the orifice diameter, $d$ is the characteristic grain size, $\rho_b$ is the mass per unit volume in 3D or per unit area in 2D of granular medium in the bulk (non-accelerating region) of the silo flow, and $k$ is a dimensionless fit parameter that accounts for steric limitations of grain motion near the orifice edges~\cite{Nedderman1992StaticsAndKinematicsGM}. Granular silo flow, in contrast with inviscid fluid flow, does not depend on the filling-height of grains in the silo~\cite{Nedderman1982DischargeRatesFromHoppers}. This empirical relation had long been derived heuristically from dimensional analysis~\cite{Beverloo1961FlowRate} and unconfirmed notions of a ``free-fall arch'' (providing the $\sqrt{g}$ factor) in the vicinity of the orifice~\cite{Tighe2007HagenTranslation,BrownRichards1970PrinciplesPowderMechanics,RubioLargo2015FreeFallArchParadox}. Very recently, however, a work-energy model~\cite{Madrid2018MuIDenseShearSilo} was successfully applied to model silo flow. By accounting for dissipation in the bulk as a dense shear flow, this model theoretically predicts both Beverloo's rule as well as the value of $C$ oft-measured for many real 3D  granular systems~\cite{Darias2020EnergyBalanceBeverloo}:
\begin{equation}\label{eqn:c}
     C = \frac{\pi\sqrt{2}}{8} \approx 0.56.
\end{equation}
This model works remarkably well for conventional horizontal orifices, but it fails to predict the dependence of flow rate $Q$ on the tilt angle $\theta$ of a silo if one assumes that the flow depends on the horizontal projected area of the orifice, \textit{i.e.}, $C\propto \cos\theta$. Numerous experiments and simulations have demonstrated, contrary to a basic and flawed prediction of this model, that flow can occur at $\theta \geq 90^{\circ}$~\cite{Franklin1955GMFlow,Chang1991VariousHorizontalVerticalOrifices,Medina2014LateralExitHoles,Anyam2022LateralOrificeDischarge,Sheldon2010TiltedHopper}. Franklin and Johanson~\cite{Franklin1955GMFlow} have proposed an empirical linear relationship between $Q$ and $\cos\theta$ with a single fit parameter that has successfully fit experimental data~\cite{Sheldon2010TiltedHopper}; however, as with Beverloo's original rule, there is no theoretical basis for the relationship. More recently, Liu~\cite{Liu2014TheoryInclinedOrifice} proposed a model based on a separate analysis for $\theta$ above and below  $90^{\circ}$. However, there are not arguments that justify a transition in the behaviour at this angle. How \textit{should} the system be modeled, then, when the silo is tilted?

The main question we address in our work, then, is how the flow rate of grains from an orifice depends on the angle of the orifice (and silo) with respect to gravity, and how therefore the model described above can successfully predict tilted silo flow rates. We approach this goal through experiments (Sec. ~\ref{sec:experiment}) with a quasi-2D monolayer of granular materials in a silo of variable tilt angle, measuring the discharge rate $Q$ with a mass scale, the outpouring velocity of grains, and the packing fraction of grains near the orifice~\cite{Janda2012SelfSimilarDensityVelocity} with a camera. Our findings (Sec.~\ref{sec:results}) suggest that a simple modification to the horizontal projection assumption (Sec.~\ref{sec:modelrepose}) that takes into account the angle of stagnant zone piles is necessary for the model to describe both our experimental results in a quasi-2D monolayer and the results of Sheldon and Durian~\cite{Sheldon2010TiltedHopper} for 3D systems. 

\section{Experiment methods\label{sec:experiment}}

\subsection{Experimental apparatus and granular material}

The system we study is a quasi-2D rectangular hopper filled with a layer of monodisperse steel bearing balls of diameter $d=3.171\pm0.001$~mm and mass $m=0.132\pm0.004$~g (uncertainties are standard deviations for over 50 measurements). 
The hopper is $75d$ wide to prevent boundary effects~\cite{Mankoc2007FlowRateGMSilo} and $150d$ high to ensure that measurable steady flow is observed in our system (see Sec.~\ref{sec:results}). There are two orifices, each equidistant from a lower corner of the hopper; we use one orifice or the other for a given angle $\theta$ such that the height of grains above the center of the orifice, in the initially full silo, is maximized. With this geometry, then, the bottom orifice is used for $\theta\leq26^{\circ}$ and the side orifice for $\theta>26^{\circ}$. We tested and confirm that choice of side or bottom orifice does not influence the results we present.

\begin{figure}
    \centering
    \includegraphics[width=0.7\linewidth]{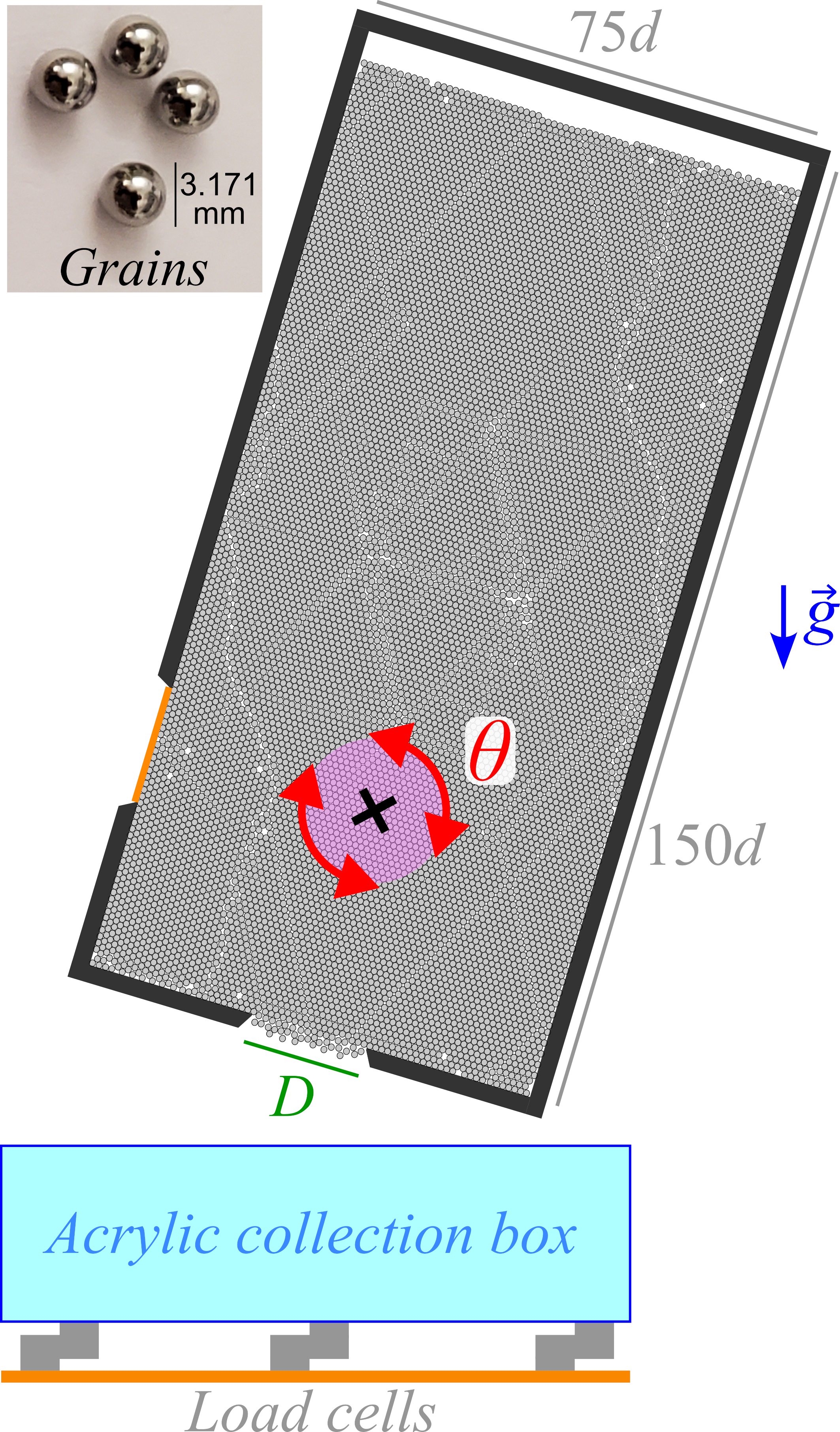} 
    \caption{Photograph of four grains and schematic diagram of system, with superposed photograph of system grains. A camera (not shown) is used to visualize the system head-on. The light bar on the left orifice represents a plug on the side orifice so that grains only flow through the bottom orifice. We only use one open orifice at any time, as described in Section~\ref{sec:experiment}. The curved arrows indicate the freely adjustable angle $\theta$ of the silo and $\times$ marks the axis of rotation.}
    \label{fig:hopperschematic}
\end{figure}

The grains are fit between two abrasion-resistant, 6.4~mm thick polycarbonate plates (Tuffak\textsuperscript{\tiny\textregistered}\ coated) that are separated by an average distance of $1.05d$ to ensure that grains are approximately monolayered and do not get wedged between the plates. The plates are reinforced with several movable 2.5~cm thick acrylic bars \textit{via} clamps at the system boundaries that are placed as necessary to prevent bowing of the front plate over the large system area. We show a schematic diagram of the system in Fig.~\ref{fig:hopperschematic}; the primary parameters we will vary are the orifice width $D$ and the angle of the orifice $\theta$ with respect to the horizontal (equivalent to the angle of the silo with respect to gravity $\vec{g}$). 

\subsection{Data acquisition}

The cumulative mass $M(t)$ of grains that exits the silo as a function of time is measured using a mass scale constructed of an acrylic collection box and three 5~kg load cells (ShangHJ 5~kg) located at the vertices of an isosceles triangle of 36~cm base length and 17~cm height. The load cells are controlled in Python with a PhidgetBridge 1046\_0 module and acquire mass readouts at 125 samples per second. The motion of grains is recorded using a camera (Basler acA1440-220um with Ricoh FL-CC1614-2M lens, image size 1440 pixel $\times$ 1080 pixel) set approximately 2~meters from the system, which is triggered periodically by an Arduino Uno to acquire images at a frame rate of 220~fps. The system is front-lit by a custom LED panel constructed from LED strips (Joylit 24~V, 6000~K) and a high frequency pulse width modulator (RioRand 15~A DC motor pump speed controller) to eliminate flickering at this acquisition rate. 

We investigate angles $\theta$ ranging from $0^{\circ}$ to the angle $\theta'$ at which flow ceases in increments of $10^{\circ}$ in most cases; $\theta'$ depends on orifice width $D$. The widths that we investigate are $D=[9.0,\;16.9,\;29.1]d$. Clogging~\cite{Janda2008JammingCriticalOutletSize} only becomes prominent for these widths at angles within $10^{\circ}$ of $\theta'$; clogging behavior is not in the scope of the current work. For all $\theta$ and $D$ in this work, at least three trials are performed to obtain an average and error bars, which denote the spread of the three trials. For each trial, the silo is rotated to be vertical, then filled with grains that are poured from a hopper in between the polycarbonate plates. The top is then loosely closed (not air-tight) to prevent grains from pouring out from this opening, and the silo is rotated to the appropriate angle. We have confirmed that filling the silo after tilting, rather than before, has no effect on the flow rates we observe.

\subsection{Analysis methods}

\begin{figure}
    \centering
    \includegraphics[width=1\linewidth]{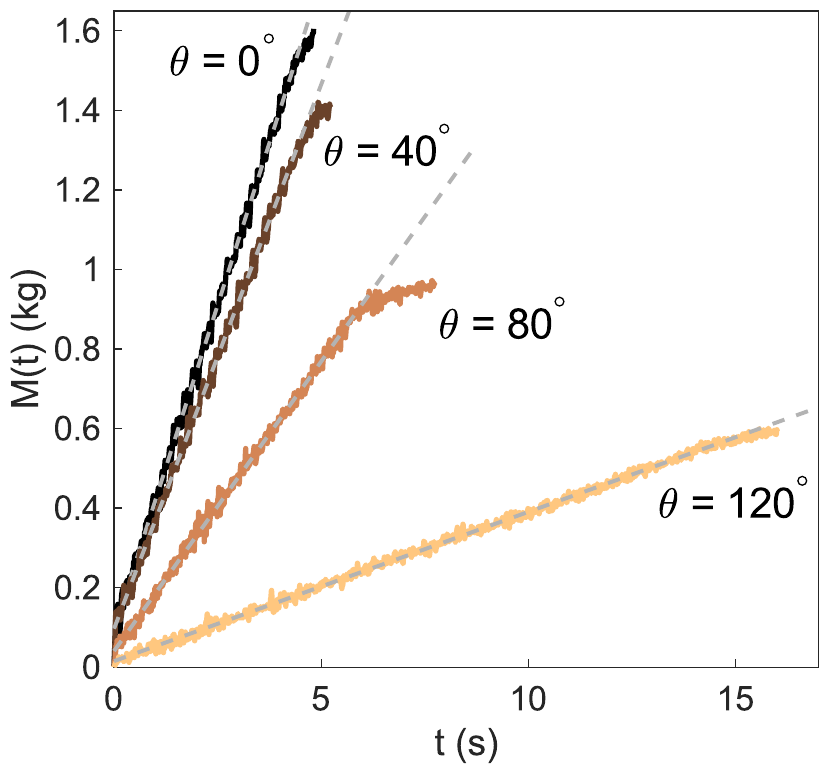}
    \caption{Example time series of $M(t)$, the cumulative mass exiting the silo, for $D=16.9d$ at tilt angles $\theta=[0^{\circ},\;40^{\circ},\;80^{\circ},\;120^{\circ}]$. Dashed gray lines correspond to best-fit lines of steady-state flow, before $M(t)$ begins leveling off. }
    \label{fig:sampletimeseriesE1}
\end{figure}

In Fig.~\ref{fig:sampletimeseriesE1}, example time series (raw data smoothed by a 5-sample wide sliding average) of the mass exiting the silo $M(t)$ are shown. The flow rate is initially steady; we fit a line to this initially linear regime to obtain the average steady-state discharge rate $Q$ for each run. The start time for the linear fit is determined by a derivative threshold crossing that indicates the start of flow (before $t=0$, not shown). The end time, beyond which the final grains dribble out at a lower rate, is picked manually at a time shortly before the slope begins to decline; we have confirmed that the slope is stable, within run-to-run fluctuations, with respect to the choice of the end time for all data sets. Note that the final asymptotic values of $M(t)$ (where the time series are truncated in Fig.~\ref{fig:sampletimeseriesE1}) decrease with increasing $\theta$ because the region of stagnant zones that remain in the silo increases as the system is tilted and fewer grains exit the silo.

\begin{figure}
    \centering
    \includegraphics[width=0.9\linewidth]{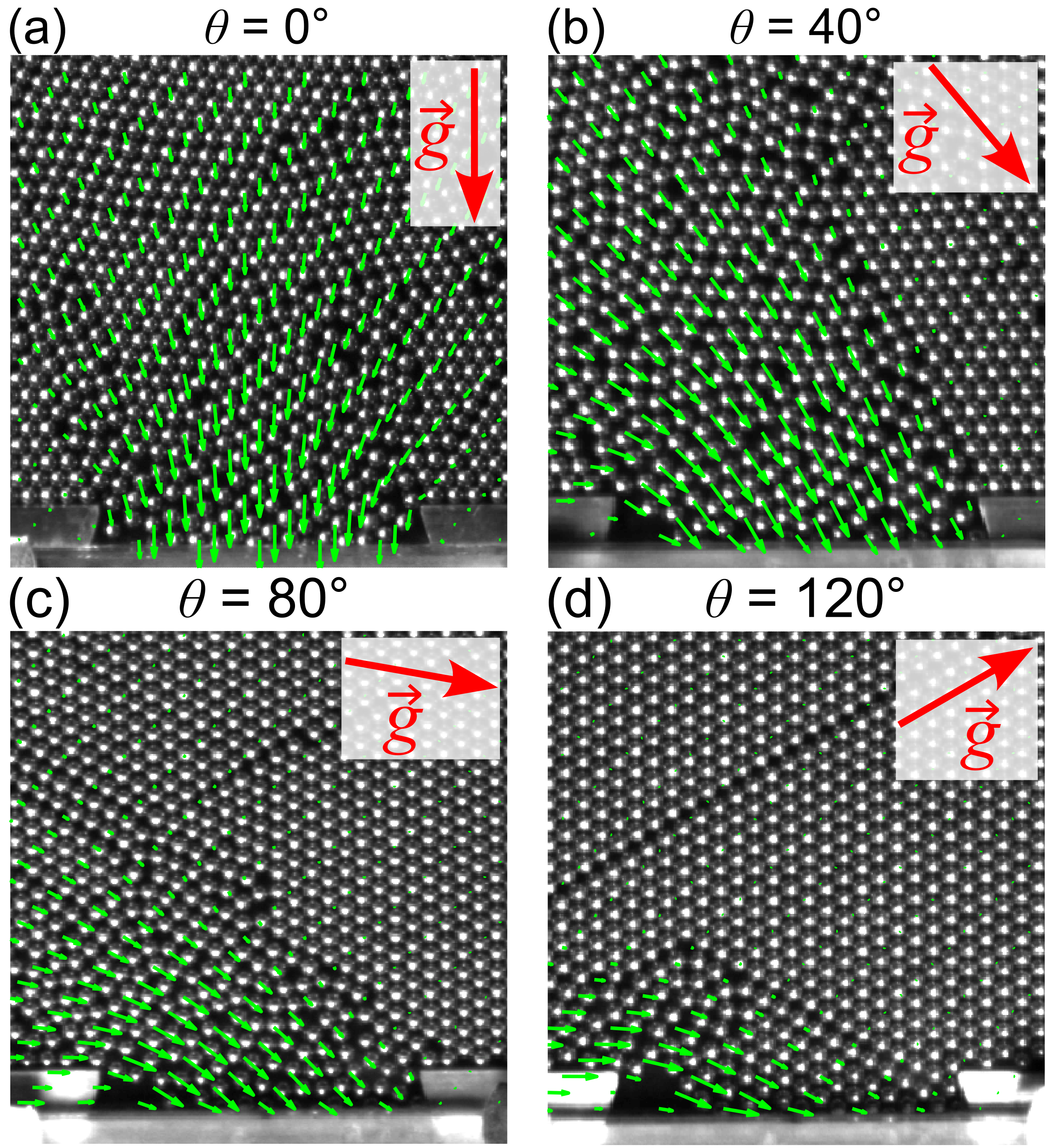}
    \caption{Example images for $D=16.9d$ at angles $\theta=[0^{\circ},\;40^{\circ},\;80^{\circ},\;120^{\circ}]$. Average PIV flow fields are shown as superposed vectors.}
    \label{fig:samplesystemimages}
\end{figure}

Example images, representative of those used to compute average flow fields and find local packing fraction near the orifice, are shown in Fig.~\ref{fig:samplesystemimages}. Corresponding videos for the sample images shown in this figure are available in the Supplementary Information. Average flow fields during the steady flow regime are computed with PIVlab in Matlab~\cite{Thielicke2014PIVlab,Thielicke2021PIVlab}, a robust particle image velocimetry (PIV) tool that searches for spatial correlations frame-to-frame to identify bulk flow patterns~\cite{Tropea2007HandbookExpFluidMech}. We have confirmed that the average flow fields we present are robust across different time ranges within the steady flow regime, and that our results are robust with the PIV settings we use. For orifice sizes $D=[9.0,\;16.9,\;29.1]d$, respectively, the first correlation search window size is $[2.3,\;3.1,\;5.4]d$ and the step size between sampling windows is $[1.1,\;1.6,\;2.7]d$. The second, more localized pass had window size $[1.1,\;1.6,\;2.7]d$ with step size $[0.57,\;0.8,\;1.4]d$. Different window sizes are needed for each $D$ because the video frame rate is fixed while the average grain speed increases as $D$ increases. 

We also compute the average two-dimensional packing fraction $\phi$, the ratio of projected particle area to available area, in the vicinity of the orifice in steady flow regime. Grains are identified in each image through straightforward morphological image operations that isolate the light glare (white spot) visible on each grain, and the area of grains occupying a region around the orifice is divided by the area of the region (see Sec.~\ref{sec:flowpattern} for more details).

Orifice outpouring velocities from PIV and $\phi$ are used to measure the flow rate from videos $Q_{video}$ as a check with $Q$ measured from the scale. With  average velocity $\vec{v}(x)$ and average packing fraction $\phi(x)$ in discrete bins of width $d/2$ along the orifice length $x=[-D/2,+D/2]$, the mass discharged per unit time is:
\begin{equation} \label{eqn:pivQ}
    Q_{video} = \sum_i \rho v_{\perp}(x_i)\phi(x_i)
\end{equation}
where $x_i$ is a discrete location along the orifice, $v_{\perp}$ is the component of velocity normal to the orifice line, and $\rho$ is the material's area density.

We note here that in Fig.~\ref{fig:samplesystemimages} and videos in Supplemental Information (and all other runs we perform) there is clear ordering of the monodisperse grains in the non-flowing regions. This ordering does not influence our results in this work; previous studies on the flow of monodisperse spheres have not shown evidence that the ordering observed affects the flow rate \cite{Mankoc2007FlowRateGMSilo}. Moreover, simulations have shown that monodisperse disks show spatial ordering but not force network ordering \cite{carlevaro2012arches,pugnaloni2016structure}, implying that contact forces, which are the ultimate responsible for the dynamics, do not show the crystal-like patterns observed in our images. We do note that simulation studies on 2D binary mixtures show a small effect on flow rate due to size dispersion \cite{zhou2015discharge,li2022influence}. However, it is unclear if this effect would be comparable at different tilt angles.

\section{Results\label{sec:results}}

\subsection{Flow rates versus tilt angle}

\begin{figure}
    \centering
    \includegraphics[width=1\linewidth]{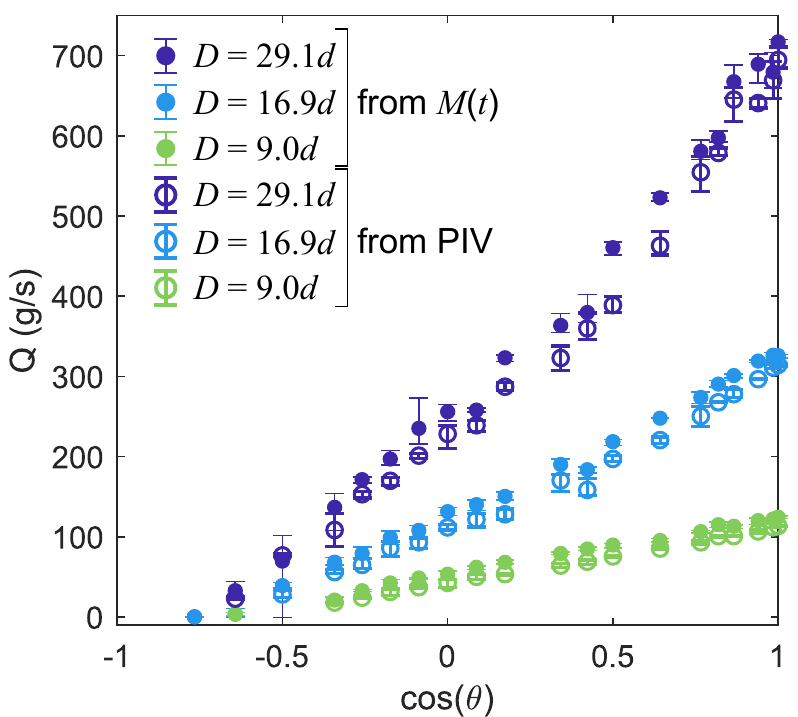}
    \caption{Flow rate $Q$ versus cosine of tilt angle for $D=[9.0,\;16.9,\;29.1]d$.  Solid symbols are for results based on $M(t)$ time series line fits (solid) as shown in Fig.~\ref{fig:sampletimeseriesE1}.  Open symbols are for results based on image analysis according to Eqn.~\ref{eqn:pivQ}. Error bars correspond to the spread of values obtained for each $\theta$ and $D$ across all experimental trials. }
    \label{fig:flowratebeforerescale}
\end{figure}

The flow rate $Q$ averaged over three trials is shown in grams/second for each width $D$ and tilt angle $\theta$ explored in our experiments in Fig.~\ref{fig:flowratebeforerescale}. The mass flow rates $Q$ were obtained from the $M(t)$ time series as well as, separately, combined information from image analysis using Eqn.~\ref{eqn:pivQ}; there is excellent agreement between these two independently measured quantities. We note here that the quantitative results we obtain in Fig.~\ref{fig:flowratebeforerescale} at $\theta=0^{\circ}$ are smaller by more than 30\% those of comparable experiments \cite{Mankoc2007FlowRateGMSilo}. However, there is significant dispersion in the literature with simulations reporting values 10\% below \cite{goldberg2015flow} and 30\% above \cite{zhou2015discharge} those reported in Ref.~\cite{Mankoc2007FlowRateGMSilo}. Since we will focus our attention on the change of $Q$ caused by changing the tilt angle, we assume that these differences affect $Q$ equally at all $\theta$.

\begin{figure}
    \centering
    \includegraphics[width=1\linewidth]{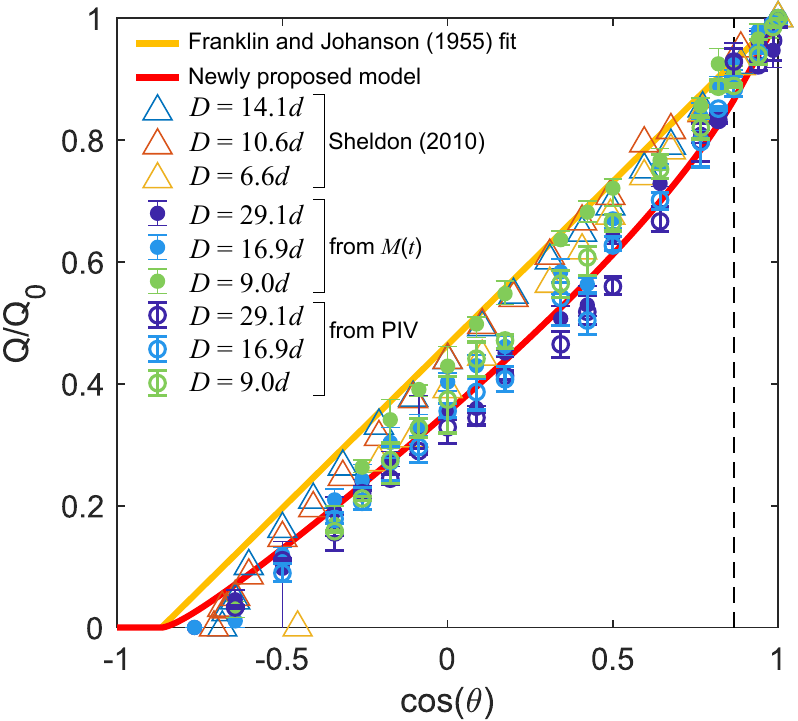}
    \caption{Normalized flow rate $Q/Q_0$ versus $\cos\theta$ for $D=[9.0,\;16.9,\;29.1]$ from $M(t)$ time series line fits (solid circles) and from PIV analysis with Eq.~\ref{eqn:pivQ} (open circles). The measurements from the 3D experiments of Ref.~\cite{Sheldon2010TiltedHopper} are included for reference (open triangles) and Franklin and Johanson's Ref.~\cite{Franklin1955GMFlow} empirical fit is included as a line. The newly proposed model in our work (Sec.~\ref{sec:modelrepose}) with $\theta_s =30^{\circ}$ is shown as a darker line.}
    \label{fig:allmassflowrates}
\end{figure}

To directly compare $Q$ as a function of $\cos\theta$ for all orifice widths $D$, we normalize by the rate $Q_0$ at zero tilt angle. 
Figure~\ref{fig:allmassflowrates} shows all data from Fig.~\ref{fig:flowratebeforerescale} with the rescaling as well as data from Sheldon and Durian~\cite{Sheldon2010TiltedHopper}. The solid lines correspond to the models described in Section \ref{sec:modelrepose}. As we can see, data from different $D$ collapse reasonably well, suggesting that the effect of the tilt angle is the same for all $D$. Likewise, the flow rate in 2D and 3D systems are equally affected by $\theta$. It is remarkable that the extrapolation $Q \to 0$ yields the same tilt ($\theta \approx 155^{\circ}$, within experimental uncertainty) for 2D and 3D systems. We will discuss in Section \ref{sec:modelrepose} that this is due to the materials and geometries used having similar stagnant zone angles next to the orifice.

We next compare our PIV analysis results with the expected flow rate given an average outpouring speed, outpouring angle, and $\theta$-independent density at the orifice. The flow rate through an orifice of width $D$ tilted by $\theta$ with average apparent density $\rho_0$ at the orifice and average outpouring velocity $\vec{v}$ across the orifice line is written as
\begin{equation}\label{eqn:q1}
    Q = \rho_0 D v(\theta) \cos[\alpha(\theta)],
\end{equation}
where $\alpha(\theta)$ is the angle between the direction of flow and the normal to the plane of the orifice. We assume that $\rho_0$ does not depend on $\theta$; we justify this assumption from our data later in this section. 

\begin{figure}
    \centering
    \includegraphics[width=1\linewidth]{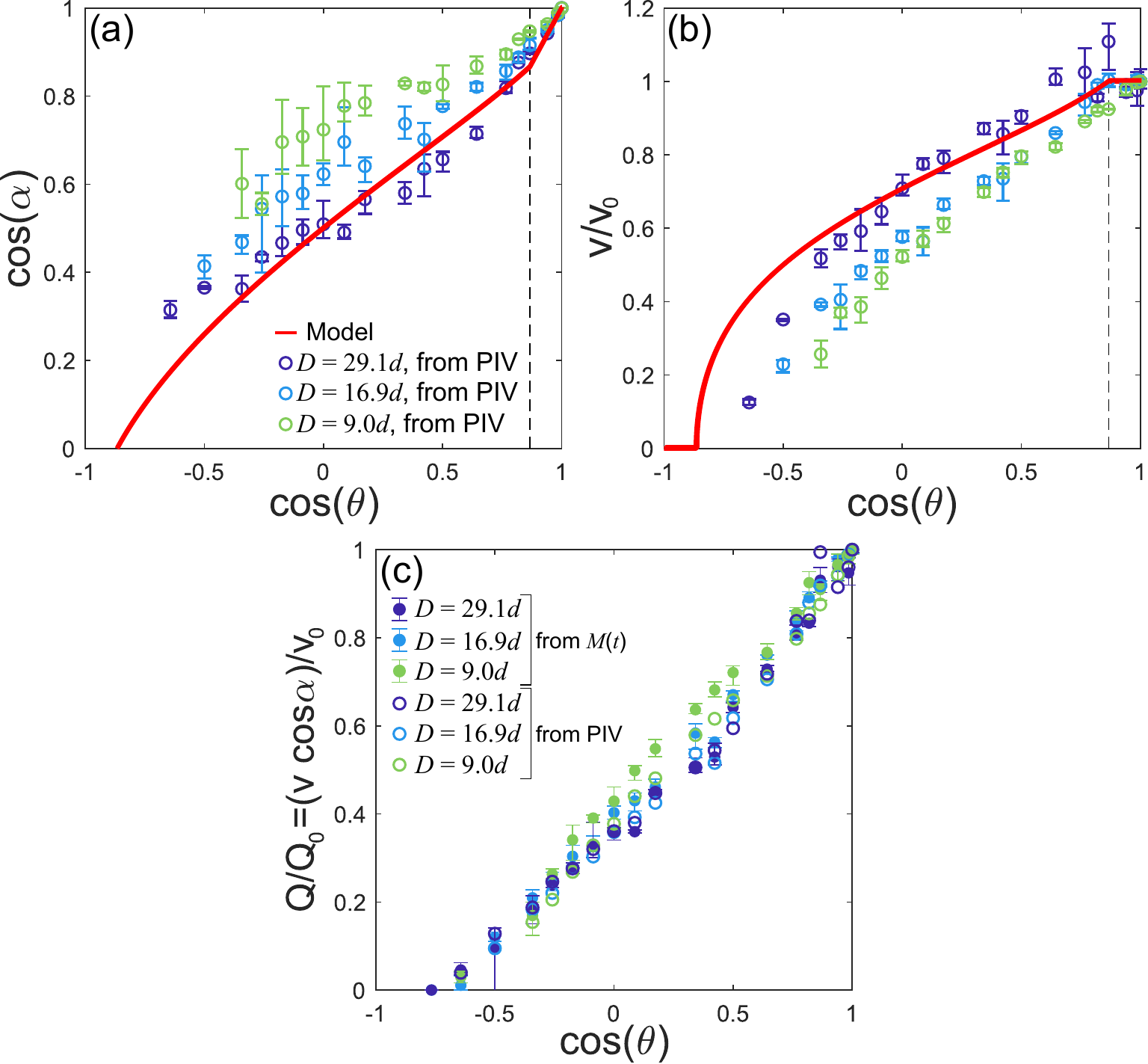}
    \caption{Comparisons of predictions from the repose model (line) and PIV analysis of particle flow fields (open circles). (a) Angle $\alpha$ between average direction of flow at the orifice and normal to the orifice plane as a function of $\theta$. (b) Normalized speed $v(\theta)/v(\theta=0) $ at the orifice as a function of $\theta$. The vertical arrow dashed line indicates the angle $\theta_s$. (c) Normalized flow rate $Q(\theta)/Q_0$ from $M(t)$ time series (solid circles) and $v(\theta) \cos[\alpha(\theta)]/v_0$ from average flow $\alpha$ and $v$ at the orifice from PIV analysis (open circles).}
    \label{fig:alpha-v}
\end{figure}

In view of Eq.~(\ref{eqn:q1}), we show in Fig.~\ref{fig:alpha-v}(a,b) $\alpha(\theta)$ and $v(\theta)$  from PIV analysis. We obtain $\alpha$ as $\alpha(\theta)=\theta-\arctan(v_y(\theta)/v_x(\theta))$, where $v_x$ and $v_y$ are the mean horizontal and vertical components of the velocity at the orifice plane. The angle $\alpha$ is zero ($\cos\alpha=1$) for the horizontal orifice ($\cos\theta=1$) and increases with tilt angle. There exists a small shift in the data for different $D$: larger orifices yield slightly larger values of $\alpha$ (smaller $\cos\alpha$). Therefore, one might infer that the flow rate drops with $\theta$ due to the reduction in the projection of the velocity vector along the normal to the orifice plane. However, Fig.~\ref{fig:alpha-v}(b) clearly demonstrates that the average outpouring speed of the particles also decreases with $\theta$. There is also here some small effect of $D$ on the speed since smaller orifices present slightly lower speeds of the outpouring particles. Overall, the effect of tilting the orifice is twofold: (i) the direction of flow and the normal to the orifice plane become non-aligned and (ii) the speed of flow decreases. The solid lines in Fig.~\ref{fig:alpha-v}(a,b) correspond to the model that will be discussed in Section \ref{sec:modelrepose} where these two effects will be taken into account.      

If the approximation in Eq.~(\ref{eqn:q1}) that $\rho_0$ can be taken as independent from $\theta$ is valid, then 
\begin{equation*}
   Q(\theta)/Q_0 = v(\theta) \cos[\alpha(\theta)]/v_0,
\end{equation*}
where $Q_0 \equiv Q(\theta=0)$ and $v_0 \equiv v(\theta=0)$. To test this, we plot in Fig.~\ref{fig:alpha-v}(c) the normalized flow rate from the data obtained using the scale along with $v(\theta) \cos[\alpha(\theta)]/v_0$ taken from Fig.~\ref{fig:alpha-v}(a,b). As we can see, the agreement is excellent indicating that $\rho_0$ does not depend on $\theta$. This finding will be further validated using image analysis in the next section.

It is interesting to note that the mild dependencies of $v(\theta)$ and $\alpha(\theta)$ on the orifice width $D$ are opposite: while $v(\theta)$ increases with $D$, $\alpha(\theta)$ decreases. As a consequence, $Q$, which depends on the product of $v$ and $\alpha$, shows only a marginal dependence on $D$ as observed in Fig.~\ref{fig:allmassflowrates}.

\subsection{Flow patterns\label{sec:flowpattern}}

\begin{figure}
    \centering
    \includegraphics[width=1\linewidth]{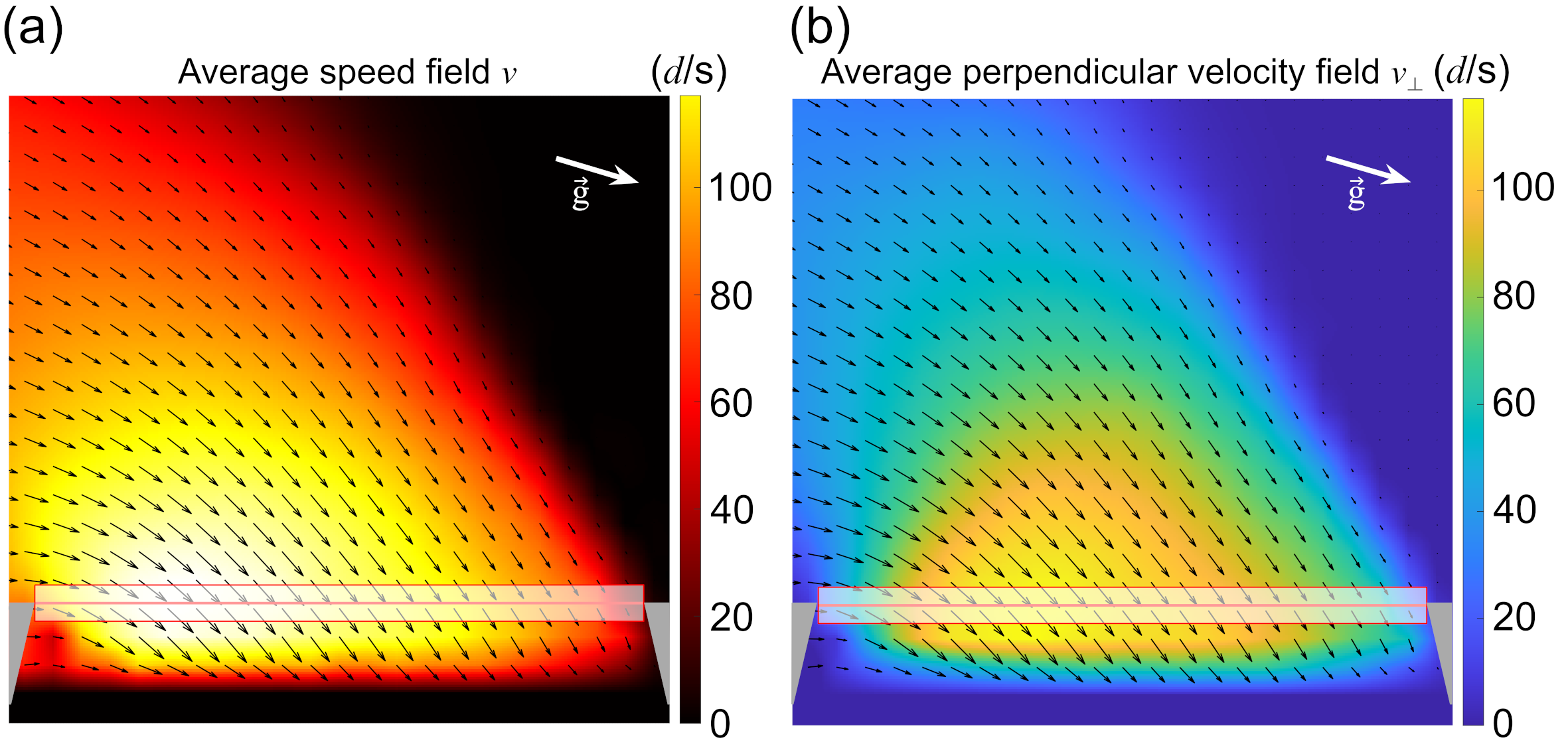}
    \caption{Heat maps for local flow speed $v$ (left) and the component of velocity normal to the orifice $v_{\perp}$ (right) for a trial with $D=16.9d$ and $\theta=70^{\circ}$. The black vectors represent the velocity field, and the white vector points in the direction of gravity $\vec{g}$. The rectangle shows the sampling region centered on the orifice (with height $d$) used to compute velocity and packing fraction profiles along the orifice.}
    \label{fig:samplecolormapspeed}
\end{figure}

\begin{figure*}
    \centering
    \includegraphics[width=1\linewidth]{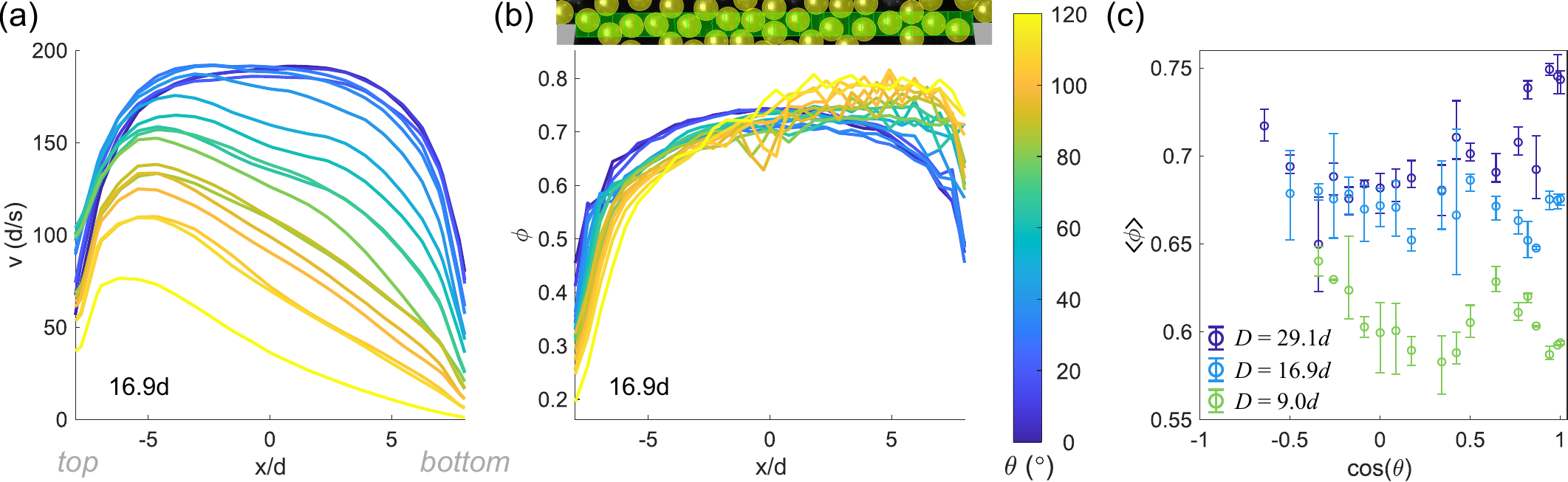}
    \caption{Average profiles along the orifice line ($x=[-D/2,+D/2]$ in units of $d$) of (a) speed $v$ and (b) packing fraction $\phi$ for different tilt angles $\theta$ and $D=16.9d$. The inset above (b) demonstrates the computation of packing fraction $\phi$ for a single frame in bins of width $d/2$ and height $d$ equally spaced along the orifice.  (c) Average packing fraction along the orifice $\langle\phi\rangle$; error bars correspond to the range of measurements for all trials. 
    }
    \label{fig:velocityprofiles}
\end{figure*}

The velocity fields from PIV and $\phi$ measurements reveal nontrivial behavior for $\vec{v}$ and $\phi$ at the orifice as the silo is tilted. Fig.~\ref{fig:samplecolormapspeed} shows the average flow field for $D=16.9d$ at $\theta=70^{\circ}$, which is clearly asymmetric with the faster flows occurring near the top (left in the figure) end of the orifice. We note here the strong resemblance between our flow fields and those shown in recent simulations for vertical orifices~\cite{Anyam2022LateralOrificeDischarge}. In Fig.~\ref{fig:velocityprofiles}(a,b) we show the profiles of speed $v$ and packing fraction $\phi$ along the orifice line, averaged over the PIV flow fields obtained from three trials in each case. As the angle $\theta$ increases (above $\sim20^{\circ}$), the flow speed decreases. Moreover, the speed approaches zero at the bottom side of the orifice while the maximum shifts towards the top side as $\theta$ increases. Correspondingly, the material becomes more densely packed near the bottom end of the orifice. We also include in Fig.~\ref{fig:velocityprofiles}(c) the average packing fraction $\langle\phi\rangle$, obtained by spatially averaging the temporal average of $\phi$ across the orifice (in bins) for each experimental trial. The packing fraction is fairly consistent across $\theta$ for each $D$ but generally is lower for smaller orifice size, when the orifice is small enough for boundary effects to be more significant.

\begin{figure}
    \centering
    \includegraphics[width=1\linewidth]{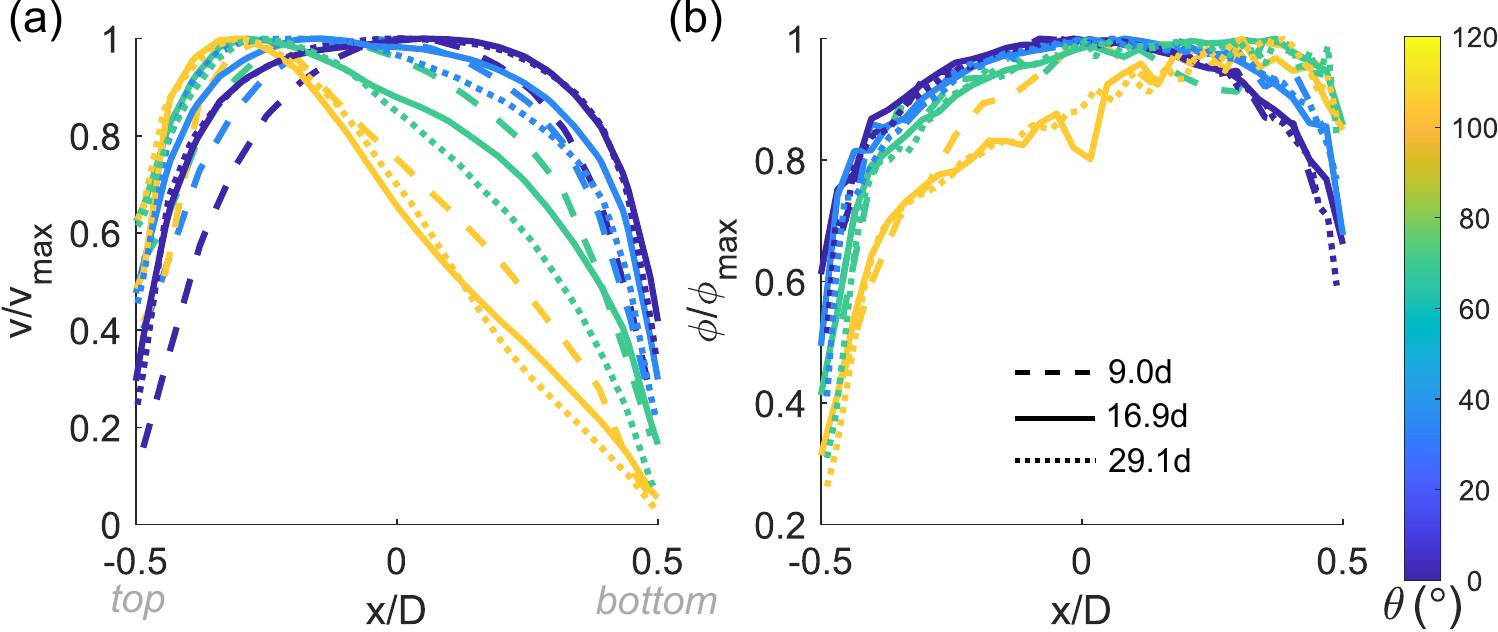}
    \caption{Average normalized profiles of (a) speed and (b) packing fraction along the orifice line for all three orifice sizes tested in this work $D=[9.0,\;16.9,\;29.1]d$ at four tilt angles $\theta=[0^{\circ},\;35^{\circ},\;70^{\circ},\;105^{\circ}]$. 
    }
    \label{fig:rescaledvelocityprofiles}
\end{figure}

Inspired by the recent work of Ref.~\cite{Janda2012SelfSimilarDensityVelocity}, in which velocity and packing fraction profiles along a non-tilted orifice are found to collapse across many orifice sizes when rescaled appropriately, we rescale the profiles in Fig.~\ref{fig:velocityprofiles}(a,b) to produce those shown in Fig.~\ref{fig:rescaledvelocityprofiles}. The horizontal axes are rescaled by orifice size $D$ to range from $-0.5$ to $+0.5$. For the speed profiles in Fig.~\ref{fig:velocityprofiles}(a), the vertical axes is scaled by the maximum speed along the orifice $v_{max}$ for each $\theta$; for the packing fraction profile in Fig.~\ref{fig:velocityprofiles}(b), the vertical axes are scaled by $\phi_{max}$, the peak $\phi$ for each profile. The speed and packing fraction along the orifice collapse well across angles $\theta$ for each orifice size examined. In agreement with Ref.~\cite{Janda2012SelfSimilarDensityVelocity}, our findings suggest that even for a tilted silo the flow near the orifice (and therefore the flow rate) is described by a universal form for any (non-clogging) orifice width $D$; a detailed study of this idea is beyond the scope of the current work, however. 

\section{Model of flow with tilt angle\label{sec:modelrepose}}

The simplest model of granular flow through an orifice tilted an angle $\theta$ assumes that the flow depends on the horizontal projected area of the orifice, \textit{i.e.}, $Q\propto \cos\theta$. While this approach is successful for small angles~\cite{Sheldon2010TiltedHopper}, it fails significantly at higher angles than about $40^{\circ}$. Sheldon and Durian~\cite{Sheldon2010TiltedHopper} found that an empirical, linear relationship $Q= A\cos\theta + B$ (proposed decades earlier by Franklin and Johanson~\cite{Franklin1955GMFlow}) captured the behavior of $Q$; this relationship has no theoretical basis, however. We instead propose a model based on the typical angle of stagnant zone regions around the orifice, where flow is mostly quiescent except for slow creep~\cite{Choi2005VelocityProfileSilosHoppers}.

\begin{figure}
    \centering
    \includegraphics[width=0.99\linewidth]{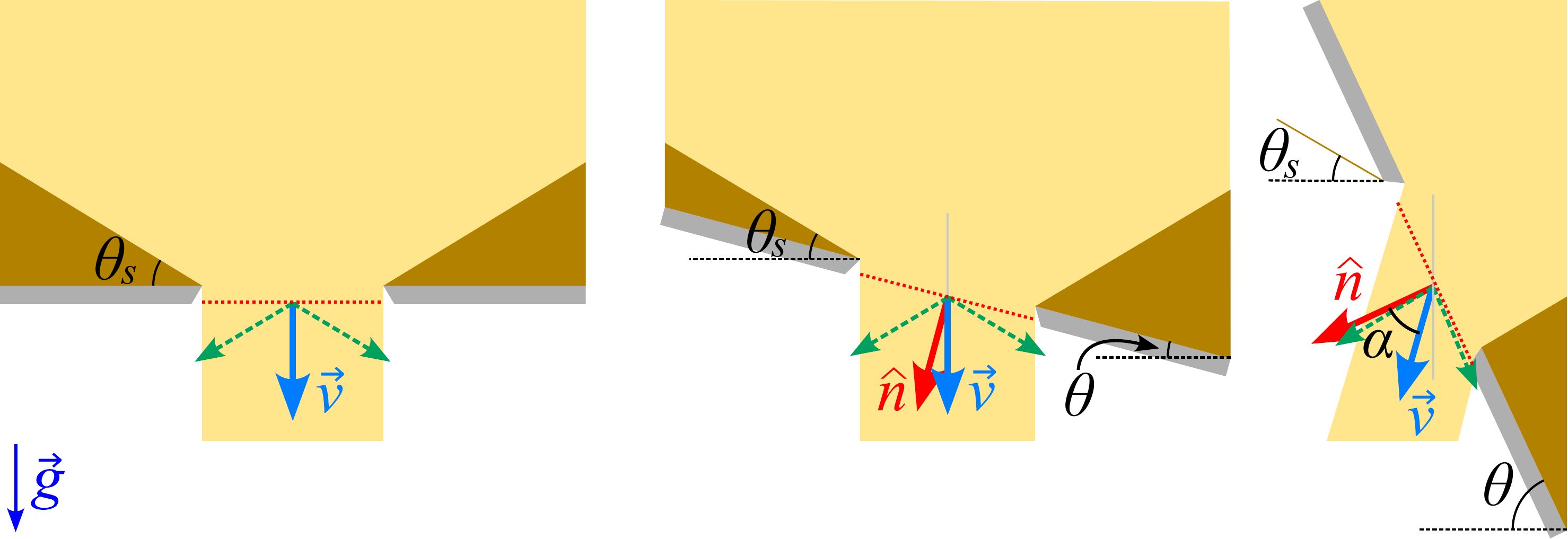}
    \caption{When the tilt angle $\theta$ is less than the angle $\theta_s$ of the stagnant zone, the side streams (dashed vectors) are symmetric and so the average outpouring velocity $\vec{v}$ is vertical and $\alpha=\theta$. However, for $\theta>\theta_s$, side streams become asymmetric as one of the orifice sides is now steeper than the angle of the stagnant zones; now $\alpha\neq\theta$.}
    \label{fig:anglereposemodel}
\end{figure}

Disregarding the non-uniform velocity and density profiles described in Section \ref{sec:results}, Eq.~(\ref{eqn:q1}) is a fair approximation for the definition of the flow rate. The correction coefficient $\cos\alpha$ in Eq.~(\ref{eqn:q1}) comes from the power extracted (from an inclined orifice) in the energy balance theory \cite{Darias2020EnergyBalanceBeverloo}. It then becomes crucial to correctly model the dependence of $\alpha$ on $\theta$. The simple assumption that the flow of grains is always in the vertical direction states that $\alpha=\theta$. Purely vertical flow, however, is only a reasonable assumption as long as there is a symmetry of outpouring grains (termed \textit{side streams} in this work), that is, when $\theta<\theta_s$, the angle of the stagnant zone pile. (Note that an angle of repose~\cite{BeakawiAlHashemi2018AngleOfReposeReview} is not well-defined in this system, so we refrain from using that term.) Figure~\ref{fig:anglereposemodel} demonstrates the importance of $\theta_s$ for the average flow direction. As the silo is tilted above the angle of the stagnant zone $\theta_s$, the flow of grains will no longer be purely vertical. As a first-order assumption, we take the average flow direction to be the bisector of the side streams along the orifice boundary. The angle between the side streams is $\pi-\theta-\theta_s$ and the  normal vector $\hat{n}$ is directed $\pi/2-\theta$ from the horizontal. It can be easily shown that $\alpha = (\theta + \theta_s)/2$. Thus we propose the following dependence of $\alpha(\theta)$:
\begin{equation}\label{eqn:cnew}
    \alpha(\theta)=\begin{cases}
			\theta & \theta \le \theta_s \\
            (\theta+\theta_s)/2 & \theta \ge \theta_s \\
		 \end{cases}
\end{equation}
Note that this is continuous at $\theta=\theta_s$.  A similar argument was invoked to model the clogging behavior of tilted hoppers (Eq.~5 and Fig.~9 of Ref~\cite{Charles2013CloggingGeometryDependence}). The solid line in Fig.~\ref{fig:alpha-v}(a) corresponds to Eq.~\ref{eqn:cnew} with $\theta_s=30^{\circ}$. As we can see, the agreement with the PIV data is fair for the largest orifice studied. Since this is a continuum model one should expect that it represents large orifices better. 

\begin{figure}
    \centering
    \includegraphics[width=1\linewidth]{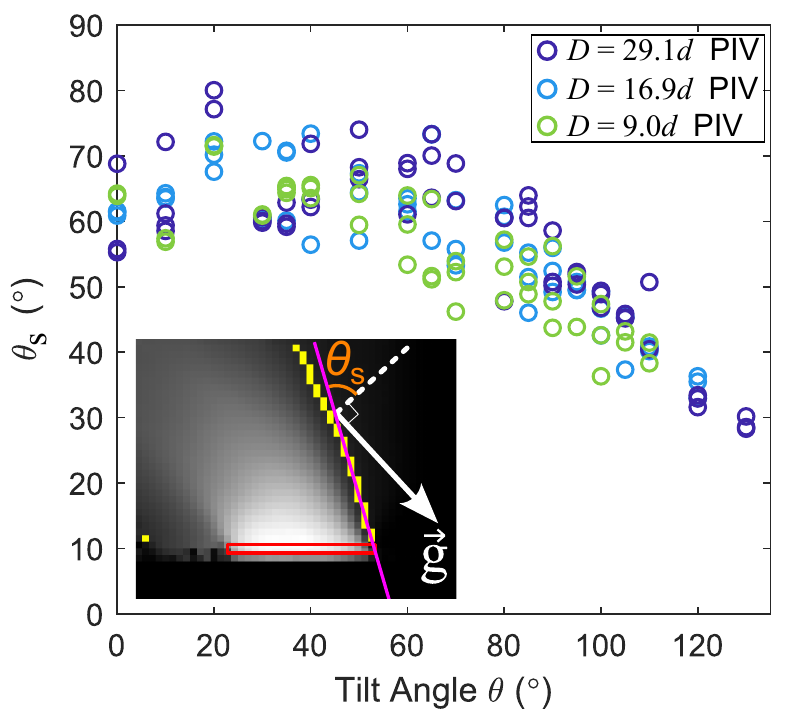}
    \caption{$\theta_s$ versus $\theta$ for all trials in this work. Inset: Speed heatmap for a trial with $D=16.9$ and $\theta = 35^{\circ}$. The fit line at stagnant zone boundary (bright pixels) at the bottom, or lower end, of the orifice (indicated by a rectangular outline) has an angle $\theta_s$ with respect to the horizontal.  }
    \label{fig:thetasvstheta}
\end{figure}

We remark here that the angle of the stagnant zone pile does vary in our monodisperse system with system preparation and tilt angle $\theta$ despite ordering. Figure~\ref{fig:thetasvstheta} shows measurements of $\theta_s$ for all three orifice sizes; the inset demonstrates how $\theta_s$ was extracted for a sample data set. The scalar speed field is thresholded ($v_{thresh} = 0.10$~m/s) to delineate a quasi-quiescent region on the bottom (lower) edge of the orifice. A line is then fit to the boundary between flow and minimal-flow regions to obtain a slope and stagnant zone angle. $\theta_s$ varies from trial to trial and below $\theta=60^{\circ}$ has a broad spread from $55^{\circ}$ to $80^{\circ}$; beyond $\theta\approx 60^{\circ}$, $\theta_s$ decreases and reaches as low as $30^{\circ}$ for the widest orifice size. Taking the largest orifice as the best large-system approximation (since our model is a continuum model), the maximum angle of flow is $130^{\circ}$ and the stagnant zone angle is a low $\theta_s\approx 30^{\circ}$, which we use as the representative value of $\theta_s$ in our model. A range of speed thresholds and methods of fitting the boundary were tested to confirm that the results in Fig.~\ref{fig:thetasvstheta} are not significantly influenced by the analysis details.

To be able to model $Q$ using Eq.~(\ref{eqn:q1}) we need an estimate for $v(\theta)$. Following the same reasoning as for $\alpha(\theta)$, we assume that $v(\theta)=v_0$ for $\theta \le \theta_s$. One can see that this is a good approximation from the velocity profiles in Fig.~\ref{fig:velocityprofiles}. For $\theta \ge \theta_s$, the flow pattern is very much affected. According to the energy balance theory \cite{Darias2020EnergyBalanceBeverloo} the velocity will depend on the square root of the characteristic length of the region of the material (above the orifice) where energy dissipation can be neglected due to the low density of the material that reduces the number of particle-particle contacts significantly. Since estimating this characteristic length from first principles is out of the scope of this work we simply propose a linear form for $v(\theta)$ that interpolates between $\theta_s$, where we assumed $v(\theta_s)=v_0$, and $\pi-\theta_s$, where the speed of flow is expected to drop to zero. Hence

\begin{equation} \label{eqn:v}
    v(\theta)=\begin{cases}
			v_0 & \theta \le \theta_s \\
            v_0\sqrt{\frac{\pi-\theta_s-\theta}{\pi-2\theta_s}} & \theta \ge \theta_s \\
		 \end{cases}    
\end{equation}

The estimate provided by Eq.~(\ref{eqn:v}) is shown as a solid line in Fig.~\ref{fig:alpha-v} with $\theta_s=30^{\circ}$. As we can see, the agreement with PIV data is fair for the largest $D$. As mentioned above, this is reasonable considering that this is a continuum approach and the effect of boundaries that are more important for small $D$ are neglected.

By plugging Eqs.~(\ref{eqn:cnew}) and (\ref{eqn:v}) into Eq.~(\ref{eqn:q1}) we obtain the flow rate that is depicted as a solid line in Fig.~\ref{fig:allmassflowrates}. The model agrees remarkably well with the experimental data. Note however that the estimates for $\alpha(\theta)$ and $v(\theta)$ in Fig.~\ref{fig:alpha-v} are less accurate. A compensation due to the underestimation of $\alpha$ and the overestimation of $v$ occurs.

As we mention in Sec.~\ref{sec:results}, $Q/Q_0$ from previous 3D experiments of Sheldon et. al. \cite{Sheldon2010TiltedHopper} agree with our 2D data. From the model we can provide a lowest-order explanation for this dimension-independent property. In 3D, Eq.~(\ref{eqn:q1}) holds by simply replacing $D$ for $D^2$, which applies equally to $Q_0$. Similarly, Eq.~(\ref{eqn:cnew}) for $\alpha$ is expected to hold with the exception of a possible difference in the value of the angle $\theta_s$. This is equally valid for the interpolation provided in Eq.~(\ref{eqn:v}) since the square root dependency in the energy balance theory is not affected by dimensionality. Therefore, unless a different angle of the stagnant zone develops, $Q/Q_0$ should not be affected by the system dimensionality. For the data of Sheldon et al., $\theta_s$ was $24^{\circ}$, which is not far from our $30^{\circ}$. It is important to mention that small changes in angle of repose are usually considered to be signals of significant differences in the properties of a granular material. However, in terms of flow rate, material properties are unimportant in a wide range of materials as demonstrated by the Beverloo's equation being valid for most materials \cite{Nedderman1992StaticsAndKinematicsGM} and explained by the energy balance model \cite{Darias2020EnergyBalanceBeverloo}.

\section{Conclusions\label{sec:conclusions}}

We have presented experimental measurements of the flow during the discharge of grains from an inclined quasi-2D silo. Besides obtaining the flow rate by weight, we also used PIV analyses to obtain flow fields and extract the mean flow speed and mean flow direction as well as the detailed velocity and density profiles across the orifice line. We have shown that the decrease of the flow rate with tilt angle is partially caused by the non-alignment of the flow direction and the orifice plane normal and partially by a decrease of the flow speed. We have provided simple models for these two effects, which allow prediction of the flow rate to first order. The only parameter in the model is the angle $\theta_s$ of the stagnant zone which, instead of fitting, we have obtained from the videos of the experiment.

It is important to emphasize that $\theta_s$ is not unique and depends on the tilt angle $\theta$, at least in this quasi-2D monodisperse system. Since our simple model considers only one single value of $\theta_s$ we have chosen the one that corresponds to the maximum tilt. It would be interesting to explore systems where $\theta_s$ can be controlled, for example by using different boundary conditions (such as highly frictional bottom boundaries), varying the dispersity of grain sizes, or modifying the preparation protocol for preparing the initial granular packing.

Finally, a more in-depth analysis of the velocity vector fields (or even rote particle tracking) becomes valuable to understand the mechanisms through which the flow speed is controlled by the inclination of the silo and orifice beyond the mean field model presented here. Future works might aim to characterize and model the profiles of velocity and packing fraction along the orifice, analyze the spatial structure of the accelerated zone in the vicinity of the orifice, and explore mechanisms of clog formation at tilt angles close to zero flow with different orifice sizes.  

\begin{acknowledgements}
RK thanks Tyler Maxwell at Berea College for construction of the apparatus and for thoughtful conversations. JCL, EO, and RK acknowledge funding for this project from Berea College's Undergraduate Research and Creative Projects Program, Summer 2022. LAP acknowledges funding from CONICET (Argentina) through grant PIP-717.  DJD acknowledges funding from NSF grant DMR-1720530.
\end{acknowledgements}

\section*{Declarations}
The authors have no conflicts of interest to declare.

\bibliographystyle{spphys}  
\bibliography{bibliography}

\end{document}